\documentclass[11pt, a4paper]{article}
\usepackage[utf8]{inputenc}
\usepackage[T1]{fontenc}
\usepackage{amsmath}
\usepackage{amsfonts}
\usepackage{amssymb}
\usepackage{graphicx}
\usepackage{fullpage}
\usepackage{multicol}

\begin{document}
\title{Quantum geometric phase with initial state index  dependent on space-time Curvature}	
\author{Jose A. Pereira Frugone \\ japfrugone@yahoo.com}

\maketitle

\begin{multicols}{2}
\textbf{Abstract} \\
We introduce a new way of generating the quantum geometric phase by making the initial base state index dependent on space-time curvature. We prove that the resulting Schr\"odinger equation is identical to the trace form of the Einstein gravitational field equations when we are in the adiabatic approximation.  This particular initial condition was proposed in a  previous work by the author.

\section{Introduction}

The quantum geometric phase arises in several contexts in Quantum Mechanics. First introduced by Pancharatnam \cite{mendas1997pancharatnam} and Berry for systems with a Hamiltonian $ H( \gamma (t)) $ dependent on a time dependent parameter $ \gamma (t) $ \cite{berry1984quantal} under the adiabatic approximation (adiabatic meaning small response of the quantum system to changes in the parameter $  \gamma(t) $). According to the state reduction postulate of Quantum Mechanics (QM) after a punctual measurement on the the quantum system at time $ t_0 $ ) we will be left with the system in one of the base states of its Hilbert space ($ |\phi_j (r,t_0) > $). The evolution of of the system state under the Schr\"odinger equation 

\begin{equation}\label{Eq1}
\hat{H} (R(t)) ~|\Phi (r,t) > = i \hbar \dfrac{d}{dt}|\Phi (r,t) >
\end{equation} 
 
and under the adiabatic conditions tells us that the system's quantum state stays close the the initial instantaneous base state, deviating only by two phase terms

\begin{align}
|\Phi &(r,t) > = \exp [\frac{1}{i \hbar}~\int _{t_0}^t E(R(t))]  \nonumber \\ 
&\exp [~\int _{t_0}^t < \Phi_j (r,t)|\dfrac{\delta}{\delta t}|\Phi_j (r,t) >] ~|\phi_j (r,t_0)>
\label{Eq2}
\end{align}

the first phase term is the usual dynamical phase term integrating over the instantaneous Energy eigenvalue of the Hamiltonian operator. The second phase term is the Geometric phase.This phase is not a physically observable quantity unless the time evolution of the parameter $ \gamma(t) $ is periodic. For periodic  $ \gamma(t) $ this phase term cannot be gauged away and becomes an observable quantity dependent on the curvature of the Hilbert space of the quantum system. \\ \\

 Other systems in which the geometric phase appears are quantum systems driven by a classical object performing periodic movement in its phase space. Famous examples are the Fermi accelerator, atoms enclosed in vibrating crystal inclusions, or the Fermi-Ulam bouncer model \cite{berman2005fermi} \cite{mostafazadeh1997eigenvalue} \cite{vubangsi2015wave}. In those systems the quantum sub-system can have holonomic boundary conditions with the classical driving system or non-holonomic ones when there is feedback between the driving system and the driven quantum sub-system. In non-holonomic situations the geometric phase induces the return of the quantum state to the initial state after some time, a phenomenon called quantum revivals. In those kind of systems the coefficients of the expansion on base states become dependent on the driving' system trajectory in phase space. \\ \\
 
 There is yet another way to get a geometrical phase in a quantum system and that is when the quantum base states of the Hilbert space ($ \mathcal{H} $) are dependent themselves on a time-depending parameter (but not otherwise  explicitly depending of time)
 
 \begin{equation}\label{Eq3}
 \mathcal{H} = \{ span~ of~|\Phi_j (\gamma (t)) >, j=1,... \}
 \end{equation} 
 
 In general $ \mathcal{H} $ is infinite dimensional an a typical quantum system will have non-null coefficients in that base only for some of the base states. Important examples of that kind of systems are at the basis of Quantum Chemistry or Condensed Matter Physics where electrons are the quantum sub-system and their states are spanned in a basis of orbital states dependent on the time varying position of the atomic nucleus. Adiabatic cases are the Born-Oppenheimer approximation for a molecular wave function. Non adiabatic cases can be constructed with the Hartree-Fock anzatz for the wave function as a product of molecular orbitals. The Schr\"odinger equation on these kind of anzatz produce geometric phase terms due to the dependency of the base on the time-dependent parameter \cite{halliday2022manifold}. \\ \\

 In this work we introduce a new way to induce a geometrical phase related to a changing Hilbert space base. Instead of making the base vectors dependent on a time-varying parameter we will make the initial base state index dependent on that parameter. This should not be confused as being the same as the case where coupling to time varying boundary conditions generate the geometric phase. In that case the time dependency is transferred to the quantum system by the Hamiltonian becoming dependent on the driver system motion. Although that will induce time dependency on the coefficients over the Hilbert space base there is no action on the base state indexes. In our new case this means that after the measurement which prepares the system for unitary evolution we are not in a randomly selected base state but in a very specific one with index given by the parameter. Concretely we propose the following form for this parameter/index
 
  \begin{equation}\label{Eq4}
 \gamma (r,t) = \beta R(r,t)
 \end{equation} 
 
 where R is the space-time curvature and $ \beta = \frac{1}{K_BT} $ with $ K_B $ the Boltzmann constant  and T a temperature parameter. This is then relevant to  Hamiltonians or base vectors dependent only on curvature. Those can correspond to systems with gravity far from matter sources or far from any singularity like for example a General Relativity, almost Minkowskin vacuum. \\
 We will show that adiabatic evolution via the  Schr\"odinger equation for this special initial state can be identified with the trace of the Einstein Gravitational Field equations. \\ 
  The choice of this concrete initial state is off course a step further from the State Reduction principle of Quantum Mechanics. That axiom states that after measurement the system will be in one unique base state randomly selected among all the possible ones in the Hilbert space. In a previous work from the author \cite{frugone2024quantum} it was proposed that this axiom is only a large scale  approximation of an almost random rule for collapse:Instead to a collapse to a random base state the system falls in the base state with (energy-ordered) index $ L \mod(n) $ where L  is the spatial scale (in Plank Length units) of the measurement system and n is the number of base states exited right before measurement. This number is almost random at usual physical scales. We will show that L mod(n) can be approximated by the above mentioned parameter $ \gamma (r,t) = \beta R(r,t) $. \\ \\

 \section{Space-time curvature dependent geometric phase}
 
 Let's assume by now that our initial state after a measurement is

\begin{equation}\label{key}
|\Phi_{\gamma(r,t_0)} (r,t_0) >
\end{equation}

with $ \gamma(r,t_0)= \beta R(r,t_0) $ (and in general $ \gamma(r,t)= \beta R(r,t) $) being the index of the base state. \\

If we are in the conditions of the adiabatic theorem,  during unitary evolution, we will stay close to that initial state. The deviation will come in the form of some additional phase terms and in addition from the slow change in the index of the instantaneous state. This change in index is what makes our path to a geometric phase different to the classical ones. We can write that state then in this way

\begin{align}
|\Phi &(r,t) > = \exp [\Omega(r,t,\gamma(r,t))] ~|\Phi_{\gamma (r,t)} (r,t_0) >
\label{Eq5}
\end{align}
 
 We can find that geometric phase $ \Omega(r,t,R(r,t)) $ by requiring this state to obey the Schr\"odinger equation followed by taking the bracket with the ket $ <\Phi (r,t) | $ and integrating in time\\ \\
 ---------------------------------------------------
 
 \end{multicols}

\begin{multicols}{1}
 
 \begin{align}
 \int_{t_o}^t <\Phi_{\gamma (r,t')}|E(\gamma(r,t')) |\Phi_{\gamma (r,t')}> dt' =  i \hbar \{ \int_{t_o}^t <\Phi_{\gamma (r,t')}|& \dfrac{d \Omega(\gamma (r,t'))}{dt'} |\Phi_{\gamma (r,t')} dt' > \nonumber + \\ 
 & \int_{t_o}^t <\Phi_{\gamma (r,t')}|\dfrac{d \phi(\gamma)}{d\gamma} \dot{\gamma} |\Phi_{\gamma (r,t')} > dt'\}
 \label{Eq6}
 \end{align}
 
\end{multicols}

 \begin{multicols}{2}
 which give us the following space-time curvature dependent form of the geometric phase        
\end{multicols}

 \begin{multicols}{1}
 \begin{align}
 \Omega(\gamma (r,t)) = \frac{1}{i \hbar} \int_{t_o}^t E(\beta R) dt' - \beta~\int_{\beta R(r,t_0)}^{\beta R(r,t)} <\phi(\beta R)| \dfrac{d}{d(\beta R)} |\phi(\beta R)> dR
 \end{align}
 
\end{multicols}

 \vspace*{1.9ex}

\begin{multicols}{2}
	
\section{Quantum statistical mechanics implications of the geometric phase}

The geometric phase can be considered as a variation in entanglement entropy ($ \Delta S $) during the evolution of the adiabatic system

\begin{equation}\label{Eq7}
\Omega(\gamma (r,t)) = \Delta S
\end{equation}

We can adopt a semi-classical form of entanglement entropy variation 

\begin{equation}\label{Eq8}
\Delta S = <E> \beta+\ln(Z)
\end{equation}

with Z being the system's partition function. We can identify terms in the following way \\

\begin{align}\label{Eq9}
\ln&(Z) =  \nonumber \\ 
&- \beta~\int_{\beta R(r,t_0)}^{\beta R(r,t)} <\phi(\beta R)| \dfrac{d}{d(\beta R)} |\phi(\beta R)> dR
\end{align}

Using the known Thermodynamics relationship 

\begin{equation}\label{Eq10}
<E> = - \dfrac{d \ln(Z)}{d \beta}
\end{equation}

applying the Leibnitz rule for the derivative by $\beta$ of the integral in r.h.s. of Eq.\ref{Eq9}  we get the following expression for $ <E> $  

\newpage

\end{multicols}

\begin{multicols}{1}

 \begin{align}
<E> = & -\int_{\beta R(r,t_0)}^{\beta R(r,t)} <\phi(\beta R)| \dfrac{d}{d(\beta R)} |\phi(\beta R)> dR -\beta R <\phi(\beta R)| \dfrac{d}{d(\beta R)} |\phi(\beta R)>|_{r,t} +\nonumber \\
&\beta R(r,t_0)<\phi(\beta R)| \dfrac{d}{d(\beta R)} |\phi(\beta R)>|_{r,t_0}- \beta \int_{\beta R(r,t_0)}^{\beta R(r,t)} \dfrac{d}{d \beta}<\phi(\beta R)| \dfrac{d}{d(\beta R)} |\phi(\beta R)> dR
\label{Eq11}
\end{align}	

\end{multicols}

\begin{multicols}{2}
We think this equation is a genuine Quantum Gravity relationship (for space time regions far from any kind of charges or singularities) at least in a semi-classical context. It connects the expected value of the Energy observable to space-time curvature. In addition, it describes how the curvature of the system's Hilbert space (reflected in the geometric phase) depends on the space-time curvature. This could be applicable for example to a slowly changing General Relativity quasi-vacuum which has an smooth or almost Minkowski metric. The last term in Eq.\ref{Eq11} is an extrinsic curvature term in the Hilbert space. All the other tems represent intrinsic curvature in the Hilbert space. We will now analyze this relationship in two limit cases and show they are related to the trace of Einstein gravitational field equations.

\section{Case with null extrinsic curvature}

Let's assume a case where the system is extremely adiabatic and very close to a Minkowskin space-time. In this case the extrinsic curvature term in Eq.\ref{Eq11} is null and we can make the the following approximation

\begin{equation}\label{key}
<\phi(\beta R)| \dfrac{d}{d(\beta R)} |\phi(\beta R)> = \omega
\end{equation}

where $ \omega $ is a constant.\\ 
In this limit Eq.\ref{Eq11} becomes 

\begin{equation}\label{Eq12}
<E>=\omega~(1-\beta)\Delta R
\end{equation}

where $ \Delta R =R(r,t)-R(r,t_0)$ \\ \\
This relationship resembles the trace of the Einstein gravitational equations 

 \begin{align}
&\kappa~T =  R (1-\frac{D}{2}) + \Lambda D\nonumber \\ \nonumber \\
& \kappa = \frac{8 \pi G}{c^4}
\label{Eq13}
\end{align}	

with G the Gravitational Constant and c is the speed of light, D the dimension of space and $\Lambda$ the cosmological constant. \\
In the case without extrinsic curvature we can consider $\Lambda = 0$. In addition far from any masses we can equal the stress-energy Tensor to have a trace equal to Energy. We have then the following form for the trace of the Einstein gravitational field equations

\begin{equation}\label{Eq14}
E= (1-\frac{D}{2}) \frac{R}{\kappa}
\end{equation}

Comparing this to Eq.\ref{Eq12}  we get the following value for the constant $\omega$

\begin{equation}\label{Eq15}
\omega = \frac{(1-\frac{D}{2}) c^4}{8  \pi G (1-\beta)}
\end{equation}

This proves that for this limit and for the evolution starting with the particular initial state we have chosen  the trace of the Einstein gravitational field equations coincide with the Schr\"odinger equation.

\section{Case with not-null extrinsic curvature}
Let's now consider a limit for Eq... with a constant intrinsic curvature for the Hilbert space 

\begin{equation}\label{key}
<\phi(\beta R)| \dfrac{d}{d(\beta R)} |\phi(\beta R)> = \omega
\end{equation}

and with a non-null extrinsic curvature. We get the following form for $ <E> $ 

\end{multicols}

\begin{multicols}{1}

 \begin{align}
<E> = & \omega (1-\beta) \Delta R- \nonumber  \beta \int_{\beta R(r,t_0)}^{\beta R(r,t)} \dfrac{d}{d \beta}<\phi(\beta R)| \dfrac{d}{d(\beta R)} |\phi(\beta R)> dR
\label{Eq16}
\end{align}

\end{multicols}

\begin{multicols}{2}

comparing the extrinsic curvature term with the Cosmological constant term in Eq.\ref{Eq13}  we get this formula for the Cosmological constant term for our General Relativity quasi-vacuum system

\end{multicols}

\begin{multicols}{1}

\begin{equation}\label{Eq17}
\Lambda = -\frac{\beta 8 \pi G}{D c^4} \int_{\beta R(r,t_0)}^{\beta R(r,t)} \dfrac{d}{d \beta}<\phi(\beta R)| \dfrac{d}{d(\beta R)} |\phi(\beta R)> dR \nonumber  \\
\vspace*{1.5ex}
\end{equation}

\end{multicols}

\begin{multicols}{2}
\section{Motivation for the particular initial condition for the quantum state}
Except for the space-time curvature dependent initial state all the results shown above are a simple application of the usual quantum mechanical unitary evolution and analysis. However the choice of that particular initial state is a step beyond the state reduction state principle of standard Quantum Mechanics. In that principle the initial state prepared after some punctual measurement falls randomly in one of the possible base states of the Hilbert space. What we are saying is that after a measurement the initial state is not a random one but instead a very specific one which looks random at the scales of a real, physical measurement. In order for all the above to have some support we need to motivate this selection and more specific the value of the index as $ \beta R (r,t) $. \\
We start by mentioning that this alternative view on the state reduction postulate has been proposed by the author in a previous publication \cite{frugone2024quantum}. In that work it was proposed that after a punctual measurement on a quantum system at space-time scale L (in Planck length units) and starting with a quantum base of size n (size meaning the maximum number of exited based states), the mixed state of the system collapses to the concrete state of index L-mod(n). We can consider an order by corresponding energy for the n states in the initial base. That makes this statement physically unambiguous. The number L-mod(n) is totally deterministic but for the typical scales of L and n it looks almost random. Small changes in L or n change the resulting index in a chaotic way which looks almost random but it is in reality completely deterministic. This form of the state reduction  principle was found in the context of a modular version of Quantum Mechanics developed by the author \cite{frugone2023quantum}\cite{frugone2024quantum} named Observation Modular Quantum Mechanics. What we have to motivate is that the number L-mod(n) is the same as our choice of index $ \beta R(r,t) $. We will not be able to prove this in this work but we can consider the following motivation:\\
The number L $ \mod(n) $ is defined via the following mathematical relationship 

\begin{equation}\label{Eq18}
L - nk = L\mod(n)
\end{equation}

where k is some integer number. L is a certain metric scale in Planck length units. Thus the l.h.s. of Eq.\ref{Eq18} is a the remaining of the division of the metric scale L by scale n. Taking L and n to be certain functions on the (r,t) space, then this difference is proportional to the curvature of the space on which those functions are defined (space time curvature R). The proportionality constant depends on the two scales we are comparing. L is a macro scale, and n is in general a quantum, micro scale. In Statistical Thermodynamics the constant connecting micro scale quantities to their ensemble level counterparts is the quantity $\beta = \frac{1}{K_B T} $. \\ 
This justifies the statement that $ L  \mod(n)  = \beta R(r,t) $.

\section{Discussion and conclusions}

We have introduced a new path for generating a quantum geometric phase for a system in an adiabatic limit. We made the index of the initial base state dependent on the parameter $ \beta R(r,t) $ with $ \beta $ being the thermodynamic beta parameter (1/($ K_B T $)) and R(r,t) the local space-time curvature. By simple application of the Schr\"odinger equation and basic analysis techniques on this initial state we found this choice of initial state induces a quantum geometric phase. This geometric phase contains terms coming from both intrinsic and extrinsic curvature of the system Hilbert space. Using the evolving quantum state with that geometric phase we found the equation for the expected value for the energy observable. This equation connects Energy and space-time curvature in a way that can be compared in certain limits to the Einstein gravitational field equations, more explicitly to their trace. We find that the limit of this equation with null extrinsic curvature in the Hilbert space corresponds to the Einstein gravitational field equations in absence of a cosmological constant and masses. For the case with non-null extrinsic curvature we found an explicit form for the Cosmological Constant term. Those results rest on the choice of the particular initial state with index  $ \beta R(r,t) $. We motivated this choice though results from a previous paper \cite{frugone2024quantum}. In it, the author proposed an alternative form of the "state reduction principle" of Quantum Mechanics where after a punctual measurement on a quantum system we don't get a randomly selected base state of the system but instead a very specific one given by $ L \mod(n) $ with L being the scale (in Planck units at which we measure the system) and n is the initial size of the Hilbert space, that is, the number of exited base states before the measurement. We showed here that the the number $ L \mod(n) $ and the number $ \beta R(r,t) $ can be thought to be the same one. The fact that this concrete choice of initial index produces a Schr\"odinger equation similar to the trace of the Einstein gravitational field equation seems to indicate that that alternative form of the "state reduction" postulate contains some elements of a potential Quantum Gravity Theory. We propose that indeed, the general equation Eq.\ref{Eq11} for <E> may be considered a genuine, semi-classical approximation of a still to be found Quantum Gravity equation. It connects Energy and space-time curvature while at the same time making the curvature of the Hilbert space dependent on the space-time curvature. \\ 
A future extension of this work could be to include  masses and working with the whole Stress-Energy tensor instead of working only for a quasi-vacuum system as we did here. Using the Dirac equation for a particle with mass and spin could be another way to extend the coverage of the ideas in this work to more realistic physical systems.

\newpage

\pagebreak

 \bibliographystyle{unsrt}
 \bibliography{bibliogra1}

\end{multicols}

\end{document}